%% file: d17.tex
\documentclass{ecai} 

\usepackage{latexsym}
\usepackage{amssymb}
\usepackage{amsmath}
\usepackage{amsthm}
\usepackage{booktabs}
\usepackage{enumitem}
\usepackage{graphicx}
\usepackage{color}
\usepackage{hyperref}
\usepackage{caption}

\usepackage{float}

\newcommand{\BibTeX}{B\kern-.05em{\sc i\kern-.025em b}\kern-.08em\TeX}

\begin{document}


\begin{frontmatter}


\paperid{17} 


\title{GeoPl@ntNet: A Platform for \\
Exploring Essential Biodiversity Variables}



\author[A]{\fnms{Lukas}~\snm{Picek}}
\author[A]{\fnms{César}~\snm{Leblanc}}
\author[A]{\fnms{Alexis}~\snm{Joly}} 
\author[B]{\fnms{Pierre}~\snm{Bonnet}} 
\author[B]{\fnms{Rémi}~\snm{Palard}} 
\author[C]{\fnms{Maximilien}~\snm{Servajean}}

\address{\textsuperscript{a}INRIA, \textsuperscript{b}CIRAD, \textsuperscript{c}LIRMM}

\begin{abstract}
This paper describes \href{https://geo.plantnet.org/}{GeoPl@ntNet}, an interactive web application designed to make Essential Biodiversity Variables accessible and understandable to everyone through dynamic maps and fact sheets. Its core purpose is to allow users to explore high-resolution AI-generated maps of species distributions, habitat types, and biodiversity indicators across Europe. These maps, developed through a cascading pipeline involving convolutional neural networks and large language models, provide an intuitive yet information-rich interface to better understand biodiversity, with resolutions as precise as 50×50 meters. The website also enables exploration of specific regions, allowing users to select areas of interest on the map (e.g., urban green spaces, protected areas, or riverbanks) to view local species and their coverage. Additionally, GeoPl@ntNet generates comprehensive reports for selected regions, including insights into the number of protected species, invasive species, and endemic species.
\end{abstract}

\end{frontmatter}

\input{sec/1_introduction}
\input{sec/2_methodology}
\input{sec/3_results}
\input{sec/4_conclusion}

\newpage
\begin{ack}
The research described in this paper was funded by the European Commission through the GUARDEN (safeGUARDing biodivErsity aNd critical ecosystem services across sectors and scales) and MAMBO (Modern Approaches to the Monitoring of BiOdiversity) projects. These projects received funding from the European Union’s Horizon Europe research and innovation programme under grant agreements 101060693 and 101060639, respectively. Further models developed based on this methodology will meet the needs of the European biodiversity strategy for 2030 through those projects. They will be used in particular to enhance the biodiversity maps at the European scale. The content of this paper reflects the views only of the authors, and the European Commission cannot be held responsible for any use which may be made of the information contained therein. 

The authors are grateful to the OPAL infrastructure from Université Côte d’Azur for providing resources and support. This work was granted access to the high-performance computing resources of IDRIS (Institut du Développement et des Ressources en Informatique Scientifique) under the allocation 2023-AD010113641R1 made by GENCI (Grand Equipement National de Calcul Intensif). 
\end{ack}
\bibliography{main}

\end{document}

%% file: sec/1_introduction.tex
\section{Introduction}
\label{sec:intro}

Global changes rapidly transform ecosystems, and their local impacts are context-dependent and hard to predict \cite{joly2024lifeclef,joly2025lifeclef}. Monitoring species composition, biodiversity indicators and habitat types at high spatial resolution is crucial for understanding ecosystem responses and aiding decision-making, but it has proven to be very challenging \cite{guisan2013predicting}. Deep learning-based distribution models offer a promising venue by allowing to use high-resolution geographic predictors and remote sensing data to address sampling gaps \cite{deneu2021convolutional}.

This work aims to leverage deep learning models for the high-resolution mapping of plant species \cite{leblanc2022species}, habitat types \cite{leblanc2024deep}, and biodiversity indicators \cite{estopinan2024mapping} across Europe. The three types of maps, developed through a multimodal cascading pipeline, provide critical insights into biodiversity patterns and ecosystem dynamics. 

By leveraging an extensive dataset \cite{geoplant2024picek} and combining both surveys made by vegetation scientists \cite{braun1932plant} and observations made by citizens \cite{contini2025seatizen}, the species distribution model \cite{joly2023overview,joly2024overview} produced species maps for over 10,000 plant species. Derived biodiversity indicators offer critical information for conservation efforts (i.e., species richness, presence of endangered or invasive species, and other conservation-relevant metrics). Furthermore, the development of high-resolution maps for habitats, generated by coupling species distribution models with habitat classification models, provides a strong foundation for understanding landscape dynamics.

\begin{figure}[t]
  \centering
  
  \includegraphics[width=\linewidth]{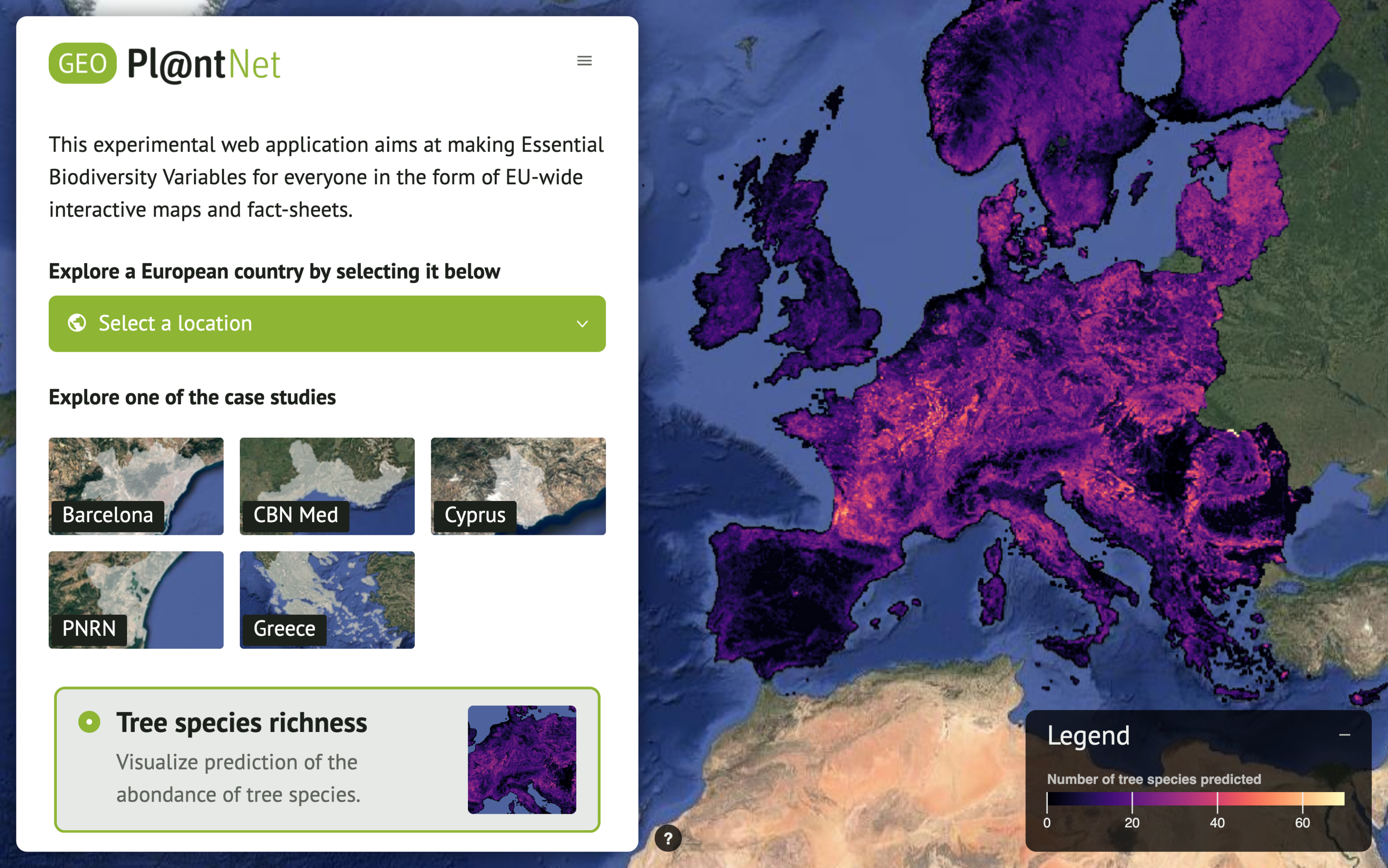}
  \captionsetup{skip=5pt}
  \caption{Home page of the GeoPl@ntNet web application.}
  \vspace{0.75cm}
  \label{fig:richness}
\end{figure}

In summary, this work represents a major step forward in producing high-resolution and large-scale biodiversity maps to support conservation and land-use planning \cite{bellard2012impacts,elith2009species}. Making them openly accessible and highly interactive through the GeoPl@ntNet web application empowers decision-makers and practitioners (see Figure~\ref{fig:richness}).

%% file: sec/2_methodology.tex
\begin{figure*}[t]
  \centering
  \includegraphics[width=0.975\linewidth]{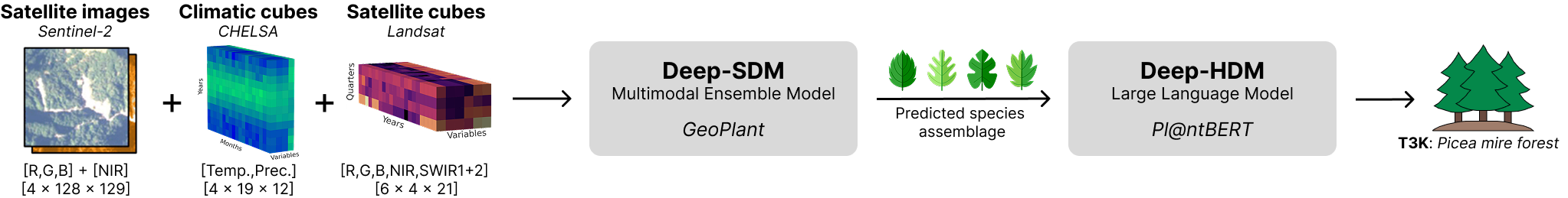}
  \vspace{4mm}
  \caption{Multimodal cascading pipeline to produce the maps. The \textbf{Deep‐SDM} (ResNet-6 ensemble) predicts species based on satellite, climatic, and environmental data, and the \textbf{Deep-HDM} (Pl@ntBERT encoder) predicts habitat types using the predicted species assemblages.}
  \vspace{6mm}
  \label{fig:pipeline}
\end{figure*}

\section{Methodology}
\label{sec:methodology}

The whole cascading pipeline \cite{leblanc2025mapping} is illustrated in Figure~\ref{fig:pipeline}.\\

\noindent\textbf{Dataset.} We use GeoPlant \cite{geoplant2024picek,botella2023geolifeclef}, covering most of European flora by aggregating 5M presence‐only \cite{bonnet2023synergizing} and 90K presence‐absence \cite{chytry2016european} records. For each 50$\times$50m cell (5.5B in total), we use Sentinel tiles, Landsat time series, and a 20-year climate record. To reflect habitat suitability, no coordinates are used \cite{cole2023spatial}. We merge data into binary occupancy labels and use target‐group background \cite{barber2022target,phillips2009sample} for bias. \\

\noindent\textbf{Species.} We use a multimodal ensemble, which outperforms classical SDMs \cite{botella2023overview,picek2024overview}. We use three ResNet-like encoders \cite{he2016deep} to process modalities independently. Embeddings are concatenated and fed to a sigmoid‐activated classifier. Training uses SGD with BCE loss. To produce maps, Europe was divided into 25×25km meta-tiles and we used the year 2021 (environmental data was averaged between March 21 and December 1, 2021), snapping water-centered cells to the nearest land. Probabilities were thresholded to yield binary maps. \\

\noindent\textbf{Indicators.} Biodiversity indicators \cite{butchart2010global} are extracted from the predicted species assemblages. They summarize ecological properties such as species richness \cite{gotelli2001quantifying}. To derive these assemblages, the probabilities are thresholded using a conformal prediction approach \cite{fontana2023conformal}, ensuring a low probability of omitting truly present species, even if it results in some false positives. This strategy prioritizes minimizing omission errors. Seven indicators (i.e., species richness, EU directive species, threatened species, most threatened, tree species, invasive species, and specialist species) have been successfully produced.

\noindent\textbf{Habitats.} We use Pl@ntBERT \cite{leblanc2024pl,leblanc2025learning}, a domain-specific language model \cite{marcos2025fully}, to assign habitat types to predicted assemblages. Pl@ntBERT performs multi-class classification into EUNIS habitat types \cite{chytry2020eunis}, leveraging its ability to capture co-occurrence patterns. The model operates purely on species presence data (i.e., no environmental variables), making it robust to sampling bias and suitable for generalization across regions. The model predicts EUNIS Level 3 habitats directly, while Levels 1 and 2 are derived from the classification hierarchy. In total, over 200 Level 3 habitat maps have been generated (keeping only the most likely label for each pixel).\\

\noindent\textbf{Website.} The front-end is developed as a single-page application using VueJS. It uses the Leaflet library to display the maps, all available as WMS streams, the standard communication protocol of the Open GeoSpatial Consortium. Thus, they can easily be integrated into any other cartographic tool. As the amount of data to be stored is huge ($\approx$15TB), an NFS mounting point is used on a Storage Virtual Machine, which we also use to store cached raster tiles provided by the MapProxy cache server and COGs used by TiTiler. The website is not just for visualization but also to get insights about any Area of Interest and export them in JSON-like format (see Figure~\ref{fig:selection}).

%% file: sec/3_results.tex
\section{Use cases}
\label{sec:results}
GeoPl@ntNet is designed to support a wide range of biodiversity-related use cases across Europe by offering an intuitive and interactive interface for exploring spatial data. It features four main modules (see Figure~\ref{fig:pages}), \textbf{Home}, \textbf{Species}, \textbf{Habitats}, and \textbf{Indicators}, each tailored to help users gain insights from high-resolution biodiversity maps. Whether the user is a conservation practitioner, urban planner, researcher, policymaker, or educator, the application facilitates exploration, analysis, and reporting with no technical overhead. A demonstration video is available on \href{https://youtu.be/secJCzVJbmw}{YouTube}.


\begin{figure}[!b]
  \centering
  \includegraphics[width=0.975\linewidth]{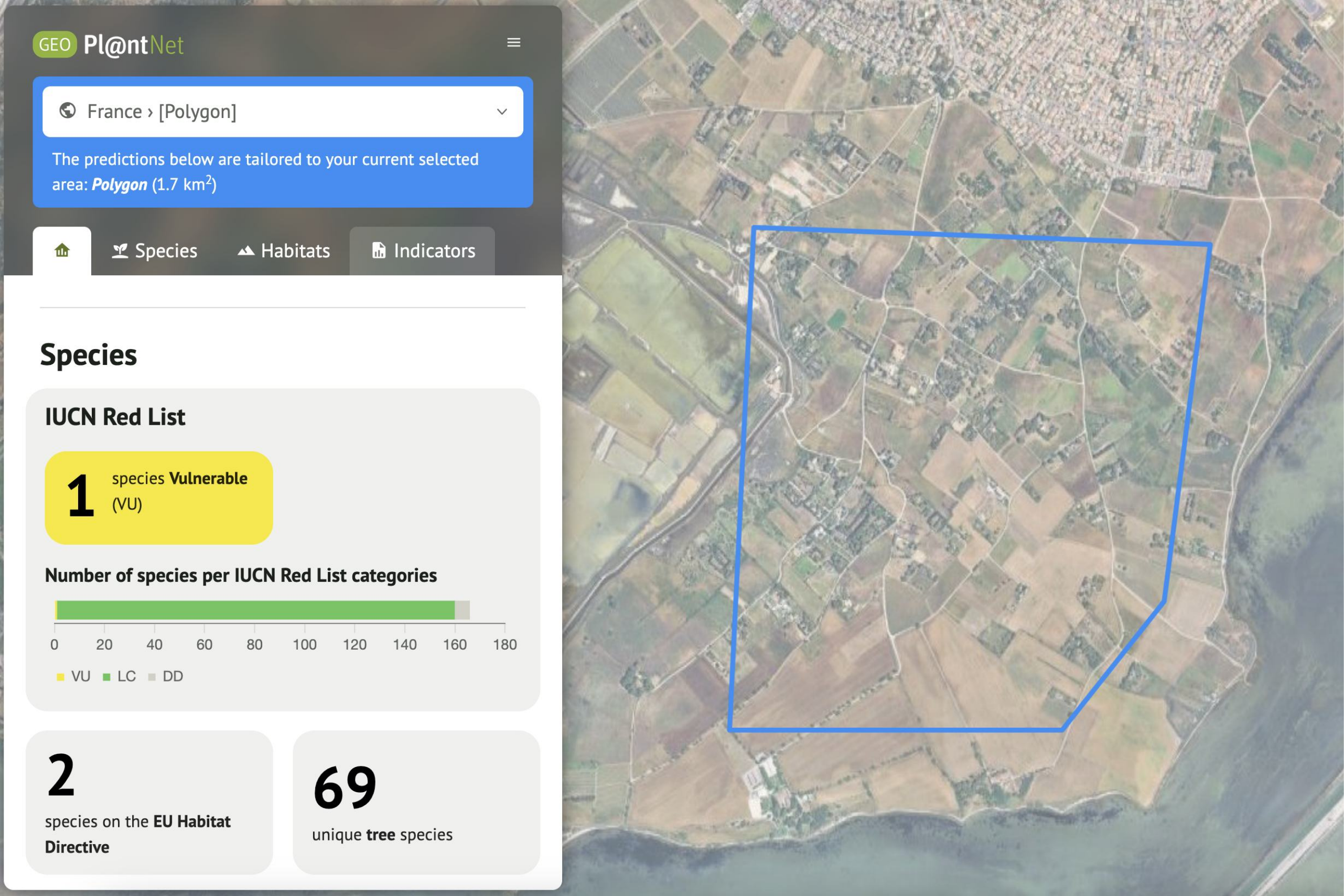}
  \captionsetup{skip=5pt}
  \caption{Area of Interest report. Users select an area and see insights on threatened or invasive species, habitat status, and other metrics.\\}
  \label{fig:selection}
\end{figure}

\newpage

\noindent\textbf{Home module.}
This module offers a comprehensive overview of the selected area, aggregating insights from the three other modules. It includes a summary of the most common and conservation-relevant species, dominant habitats, and biodiversity scores. It is the starting point for users who want a quick, readable biodiversity report. \\

\noindent\textbf{Species module.}  
This module allows users to explore the predicted distribution of plant species within any selected area. After choosing a region of interest, users are presented with a ranked list of species likely to be found there (see Table~\ref{tab:deep-sdm}). Each species entry is enriched with coverage statistics and taxonomic information, helping users assess which species are most representative or dominant in the area. The tool supports use cases such as drafting species inventories, identifying candidate species for restoration projects, and detecting potential newcomers in changing environments. \\

\noindent\textbf{Indicators module.} 
This module provides a snapshot of ecological health through a set of biodiversity indicators computed from the local species assemblage. Users can instantly see how an area scores in terms of different metrics. This functionality supports monitoring programs and conservation prioritization. For instance, users can quickly compare areas based on their conservation value, while NGOs can identify high-priority zones for habitat protection. \\

\noindent\textbf{Habitat module.}
This module allows users to visualize the most probable habitat types in the selected region. It allows filtering and sorting by habitat level and coverage, enabling a focused inspection of habitat diversity and dominance. Users can explore the landscape composition, change hierarchical levels (see Table~\ref{tab:plantbert}), and download information relevant for land-use planning, environmental education, or ecological research. This is especially useful for assessing the ecological impact of land transformation. \\

Across all modules, users benefit from a seamless map-based interface where they can pan, zoom, and draw custom regions of interest. Reports are generated in real time, allowing for rapid exploration and comparison of multiple areas (see Figure~\ref{fig:example_maps_all}). GeoPl@ntNet is not just a visualization tool, it is a decision-support system that turns large-scale biodiversity predictions into accessible insights. \vspace{-1mm}

%% file: sec/4_conclusion.tex
\section{Conclusion}
\label{sec:conclusion}

This work presents \href{https://geo.plantnet.org/}{GeoPl@ntNet}, an interactive web application based on a multimodal deep learning framework. Its main principle is to enable users to explore high-resolution (50m), AI-generated, maps of plant species, habitat type, and biodiversity indicator distribution. The whole of Europe can be explored this way, at any scale and for the majority of vascular plants, seven different biodiversity indicators and all terrestrial habitats. It can also interactively explore the biodiversity in a user-defined region of interest, bridging the gap between large-scale ecological modeling and real-world decision-making.

\begin{figure*}[h]
  \centering
  \begin{minipage}[t]{0.49\textwidth}
    \captionof{table}{Evaluation of the deep-SDM (with different metrics).}
    \label{tab:deep-sdm}
    \vspace{3mm}
    \centering
    \begin{tabular}{@{}lc@{\hspace{3mm}}c@{\hspace{3mm}}c@{\hspace{3mm}}c@{}}
      \toprule
      \textbf{Branch} & \textbf{AUC} & \textbf{F\textsubscript{1}} & \textbf{Recall@50} & \textbf{Recall@250} \\
      \midrule
      All (Sen+Bio+Lan) & 0.931 & 0.338 & 0.639 & 0.908 \\
      \bottomrule\\
    \end{tabular}
  \end{minipage}
  \hfill
  \begin{minipage}[t]{0.49\textwidth}
    \captionof{table}{Evaluation of Pl@ntBERT (at different levels).}
    \label{tab:plantbert}
    \vspace{3mm}
    \centering
    \begin{tabular}{@{}lccc@{}}
      \toprule
      \textbf{Top-SDM predictions} & \textbf{Level 1} & \textbf{Level 2} & \textbf{Level 3} \\
      \midrule
      Keeping first 100 species &76.30\% & 62.68\% & 44.72\% \\
      \bottomrule\\
    \end{tabular}
  \end{minipage}
\end{figure*}

\vspace{0.5mm}

\begin{figure*}[h!]
  \centering
  \includegraphics[width=0.975\linewidth]{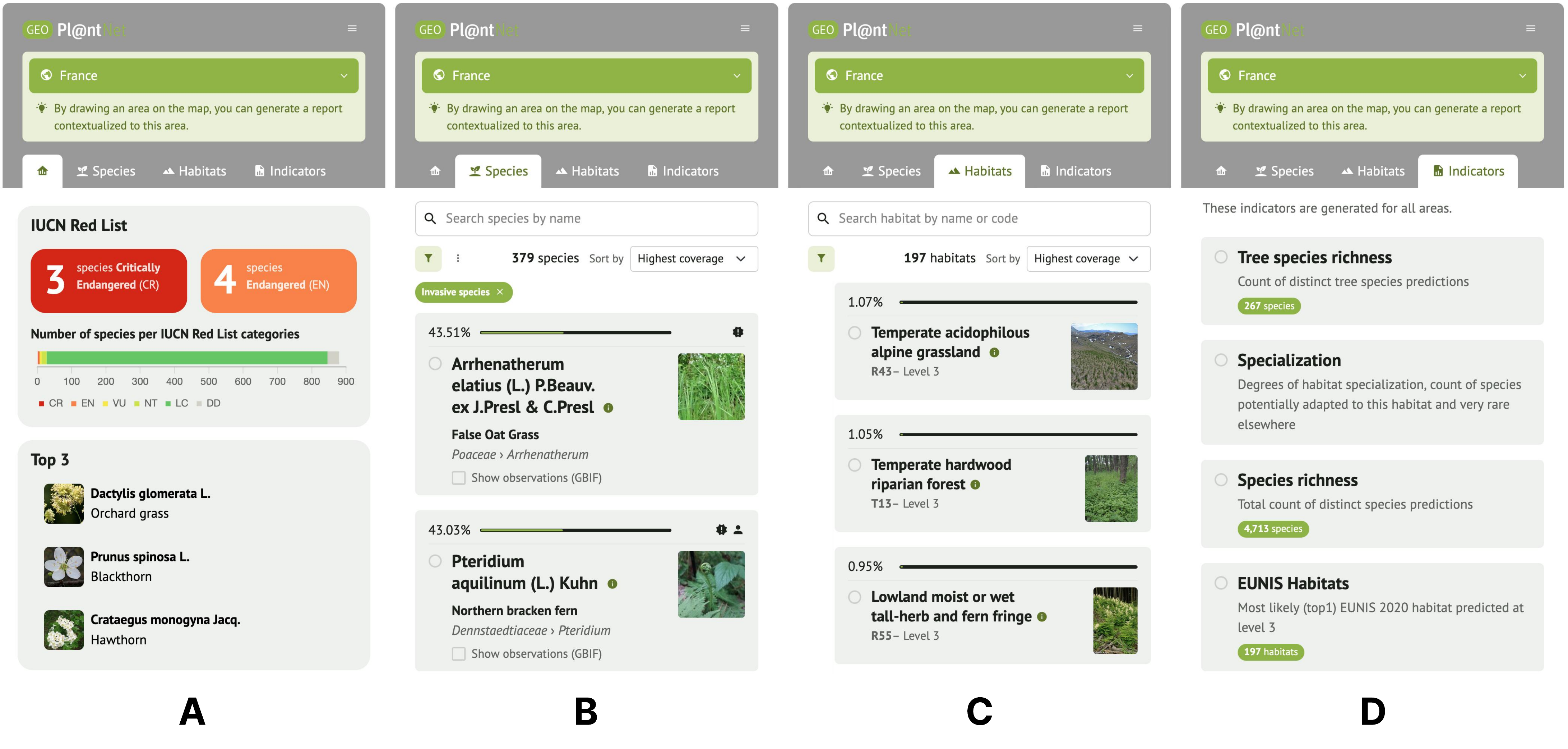}
  \captionsetup{skip=5pt}
  \caption{Modules available. \textbf{Home module} (A): Offers an overview of key biodiversity metrics. \textbf{Species module} (B): Displays a ranked list of species predicted. \textbf{Habitats module} (C): Shows the most likely habitat types. \textbf{Indicators module} (D): Summarizes biodiversity indicators.\vspace{0.5cm}\\}
  \label{fig:pages}
\end{figure*}

\begin{figure*}[h!]
  \centering
  \includegraphics[width=0.975\linewidth]{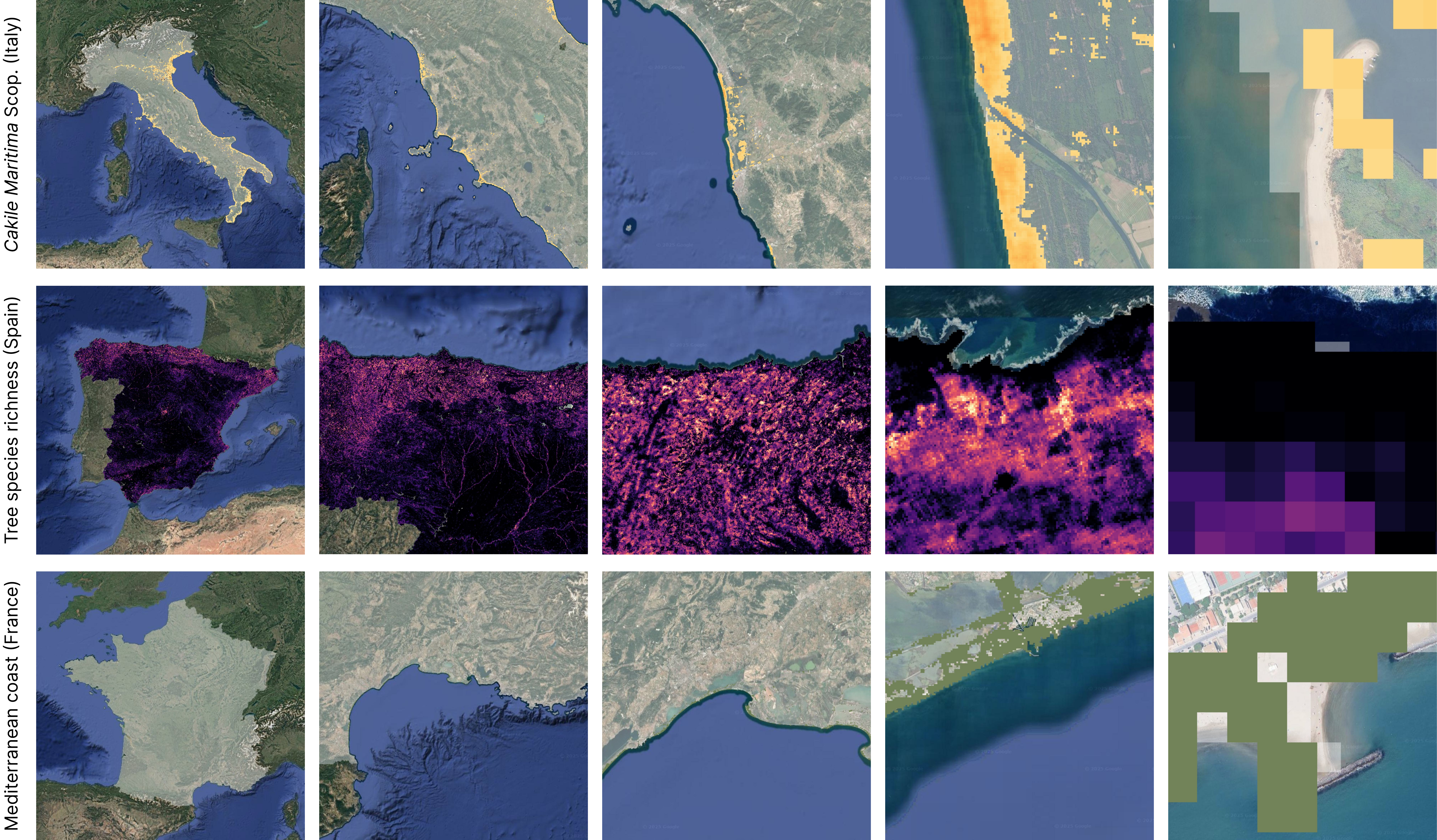}
  \captionsetup{skip=7pt}
  \caption{Example maps. Top panel shows \textbf{species distribution maps}, middle panel shows \textbf{biodiversity indicator maps}, and bottom panel shows \textbf{habitat suitability maps}. They are available at a 50m resolution in Europe for over 10,000 species, over 200 habitats, and 7 indicators.}
  \label{fig:example_maps_all}
\end{figure*}